\documentclass[
  ]
  {aipproc}

\layoutstyle{6x9}

\begin{document}

\title{Water delivery in the Early Solar System}

\classification{
                \texttt{96.12.Bc, 96.30.Ys}}

\author{Rudolf Dvorak}{
  address={Universit\"atssternwarte Wien, T\"urkenschanzstr. 17, 1180, Wien, Austria},
  email={rudolf.dvorak@univie.ac.at}
}

\author{Siegfried Eggl}{
  address={Universit\"atssternwarte Wien, T\"urkenschanzstr. 17, 1180, Wien, Austria}
}

\author{\'Aron S\"uli}{
  address={Universit\"atssternwarte Wien, T\"urkenschanzstr. 17, 1180, Wien, Austria},
  altaddress={Department of Astronomy, E\"otv\"os Lor\'and University, P\'azm\'any P\'eter s\'et\'any 1/A, 1117 Budapest},
}

\author{Zsolt S\'andor}{
  address={Universit\"atssternwarte Wien, T\"urkenschanzstr. 17, 1180, Wien, Austria}
}

\author{Mattia Galiazzo}{
  address={Universit\"atssternwarte Wien, T\"urkenschanzstr. 17, 1180, Wien, Austria}
}

\author{Elke Pilat-Lohinger}{
  address={Universit\"atssternwarte Wien, T\"urkenschanzstr. 17, 1180, Wien, Austria}
}

\begin{abstract}
As part of the
national scientific network 'Pathways to Habitable Worlds' the delivery of 
water onto terrestrial planets is a key question since water is essential for the 
development of life as we know it. After summarizing the state of the art we
show some first results of the transport of water in the early Solar System for scattered main belt objects.
Hereby we investigate the questions whether planetesimals and planetesimal fragments which have gained considerable
inclination due to the strong dynamical interactions in the main belt region around 2 AU can be efficient 
water transporting vessels. 
The Hungaria asteroid group is the best example that such scenarios are realistic. 
Assuming that the gas giants and the
terrestrial planets are already formed, we monitor the collisions of scattered 
small bodies containing water (in the order of a few percent) 
with the terrestrial planets. Thus we are able to give a first estimate 
concerning the respective contribution of such bodies to the actual water content in the crust of the Earth.
\end{abstract}

\maketitle


\section{Introduction}

The presence of liquid water on the surface of a terrestrial planet is a basic
requirement for habitability in planetary systems. 
The questions one needs to answer in this connection are

\begin{itemize}
\item When the terrestrial planets formed how much was their content of water?
\item Why don't we find water in the same quantities on the other terrestrial planets?
\item What happended to the water when a mars-sized object hit the Earth and the Moon formed?
\item What happened during the Late Heavy Bombardement (LHB)
\item Where from came water after the LHB?
\item What is the role of the comets from the Oort Cloud?
\end{itemize}

A main problem in this context is to find out where the water came from in the 
early stages on the one hand; on the other hand, when water was lost 
during special phases in later stages one needs to explain how it was 
replenished on the surface. At the end one should explain also how it could 
stay liquid on a terrestrial planet in the habitable zone for times up to 
billions of years. A central question in this respect is the collisional 
behaviour of small bodies regarding their content of water; it has to be 
modelled with specially designed effective programs like the well known 
SPH (Smooth Particle Hydrodynamics) codes. 

The possible water loss of terrestrial like planets in our Solar System (SS) and in Extrasolar Planetary system (EPS) in general should be set in context with 
geophysical processes like the stop of outgassing due to rapid mantle and core
cooling or lack of atmospheric protection by a planetary magnetosphere
\begin{itemize}
\item the stellar radiative environment of young active stars (SS and EPS) 
\item collisions of protoplanetary objects in general (SS and EPS), 
\item the Late Heavy Bombardement (SS)
\item the formation of the Moon (SS)
\end{itemize}

After the early stage the transport mechanisms in our SS from the 
main belt and also the Edgeworth-Kuiper-Belt can adequately be computed taking 
into account the important role of all sorts of resonances: mean motion
resonances, secular resonances and three body resonances. The water delivery
of the comets from the Oort cloud can be investigated statistically although 
it can only account for a fraction of the water on Earth regarding their 
different D/H ratio. Comets may have been brought into the inner SS 
by orbital changes due to passing stars, interstellar clouds and galactic
tides leading to comet showers. Although the water transport is the 
central question it is of fundamental interest to investigate how organic 
(carbon-containing) material could be delivered which then lead or may have 
lead to the development of life on Earth-like planets in habitable zones.

Different scenaria will have formed very different architectures of the
planets in an extra solar planetary system compared to our own system. 
The observed, close-in, Jupiter like planets, which evolved into such orbits 
via migration processes, make it difficult to explain the continuous existence 
of terrestrial planets on stable orbits within the habitable zone. Together 
with theoretical investigations on habitable planets, results from the 
existing satellite missions (CoRoT, KEPLER and Herschel) as well as 
future ones (Plato, James Webb, Gaia) combined with the progress in 
Earth-bound observations (Alma, ESO) will help to clarify the origin and 
presence of water (and organic materials) as a basis for life.

\section{State of the art} 

From many articles concerning the formation of terrestrial planets and their 
content of water (e.g. \cite{Lun03,Ray04,Mor00,Ali10,Ray11}) we can draw a 
coherent picture of those phases of planet-formation where the debris disk 
and a giant planet were already present. Following current models, most 
of a planet's water-content can be regarded as being produced by collisions 
between the growing protoplanet and Moon to Mars-sized planetesimals
originating from the asteroid belt. According to \cite{Lev03} and also 
\cite{Ray04} the accretion of planetary embryos from distant regions 
(outside the snowline) by terrestrial planets could have happened also without the 
presence of a Jupiter-sized object. Other studies claim that the early Earth 
as well as the terrestrial planets were dry, just as the asteroids 
in the region of their formation, because only in the cold outer part of the
early SS gas and 
water were present in big quantities (\cite{Gri11}). But in these phases 
collision events 
(\cite{Chy90a,Chy90b}) as well as the EUV radiation from the early star could 
have reduced the water content in these regions 
(e.g. \cite{Cha96a,Cha96b,Lam11}). At any rate during a later stage water 
was brought onto the surfaces of the terrestrial planets and, whereas Venus 
and Mars could not keep their water on the surface, the Earth's magnetosphere 
inticipated water loss (e.g. \cite{Lam11}). Many scenarios 
try to explain the water transport onto Earth; the most plausible seems to be that 
the C-asteroids from the outer main belt of asteroids, main belt comets (\cite{Ber11}) and small 
bodies from outer regions of the SS up to the scattered disk,
consisting in big parts of frozen water contributed to the water 
content on Earth. 

Given the discovery of water and a subsurface ice reservoir on the asteroid 
24 Themis (\cite{Cam10}), and comet-like activity of several small asteroids 
it is clear that water is in fact abundant in many solar system bodies and 
may even lie well hidden inside a crust. Collision probabilities, 
impact velocities and size distributions depend crucially on the orbits 
of the colliding objects as well as the perturbations of the planets on their 
motion respectively. Regarding these topics, namely

\begin{itemize}
\item formation and early development of Earth-like planets with respect to their water content, 
\item the possible loss of water through collisions with other celestial bodies (e.g. the impact of a Mars-like body onto the Earth with subsequent formation of the Moon, (\cite{Pah07}, \cite{Tay10}) and 
\item late water transport,
\end{itemize}

cannot be modelled by pure gravitational N-body simulations, but with sophisticated codes
including accretion, the role of the disks, the  collisional growth 
etc.(\cite{Egg10}). Numerous simulations concerning the formation of planets in the early 
Solar System have been performed where the early formation of a gas giant 
(Jupiter) is assumed. 
Hereby the giant planets are 
playing a key role; they formed when still there was a considerable 
amount of helium and hydrogen present in the early Solar nebula. 
Later accretion of terrestrial planets is closely connected to the 
perturbations due to these planets on planetesimals within the inner part 
of the disk (\cite{Ray04}, \cite{Ali10}. The process of accretion of 
embryos by terrestrial planets may be possible for different giant 
planet configurations, and even without gas giants present in the system 
(e.g. \cite{Lev03} and \cite{Ray04}). Although most of the discovered EPS 
host at least one big planet -- due to a biased sample 
because of the constraints in our observations -- this may not be the rule for 
the formation of planetary systems in general
(e.g. CoRoT-7b,c\footnote{CoRoT-7 is a planetary system (consisting of at least
two planets) which was discovered by the space mission CoRoT (details in \cite{Hat10})}). 

A crucial factor for water-delivery scenarios onto terrestrial planets 
is the so-called 'snowline' which is due to the outward diffusion of gas 
charged with vapour that condensates on existing particles during the period
when its temperature changes. 
This change acts on the accumulation of particles 
that originate from further radial distances and have a faster inward 
migration because of to their small sizes. 
As a consequence, water 
can be present as water ice bound in icy planetary embryos in the outer 
parts (respectively beyond the snowline) of the protoplanetary disc. Accretion of water from these bodies 
is a stochastic process, therefore planets may have different water content 
due to their different histories (\cite{Mor00}). In this article it is 
also claimed that 
such contributions to water on terrestrial planets may be minor because of 
the perturbations of Jupiter. There exist quantitative estimates for the 
impact erosion 
of atmospheres and condensed oceans of planets during 
the LHB (\cite{Chy90a}). But also the delivery of prebiotic 
organic matter (C, H, O, N and P) together with water by main belt comets and also 
comets from the Oort Cloud(\cite{Pie06}) has been established via hydrodynamic simulations. 
According to recent results of computations by \cite{Abr09} some small 
amount of amino acids could even survive low impact velocities as subsurface habitats.

\section{Transport of water to the terrestrial planets from the Hungaria main belt region}

One expects that the main source for water delivery to the Earth are asteroids
in the main belt between Mars and Jupiter (as well as comets from the 
Oort cloud).
A water gradient in the protoplanetary disk such that at 1 AU
bodies were dry, whereas bodies at 2.5 AU contain 5 percent of water is the
usual assumption.
It is well known (\cite{Mil10}) that asteroid groups in the main belt 
with high inclination to the ecliptic plane 
can evolve to become Mars crossers. Such configurations seem promising
candidates, if one was to look for possible mechanisms
  that can uphold a constant supply of material into the inner Solar System.

In our preliminary approach we took a sample of fictitious small bodies in the 
region where now the Hungaria family of asteroids is
located. This family is believed to originate form a violent dynamical event 
(\cite{Mil10}, \cite{War09}) about 0.5 Gyrs ago that
caused an injection of the Hungaria predecessors into orbits with an
inclination of about 20 degrees.
Another interesting point is the proximity of these asteroids to the 
4:1 mean motion resonance  with Jupiter as well as several secular 
resonances as their semimajor axes are mostly  between 1.8 and 2 AU). 
Nowadays the Hungaria group consists of more than 8000 known members with 
the largest objects with sizes up to 12 km. The membership of asteroids within
this group to one or more families is still in debate (\cite{Mil10}). 
However, for our purposes, the existence of such bodies will be taken as a 
reasonable argument, that during dynamically more violent 
times in the lates stages of the Solar System's formation,  
planetesimals could have been proliferated to this region of the main belt.

We have undertaken numerical simulations up to 40 million years, 
in order to investigate 
the number of possible close encounters respectively impacts of our test
population with the terrestrial planets in the inner Solar System. 
As a dynamical model we chose to include the Venus-Earth-Mars-Jupiter-Saturn system as it is now, 
with exception that we did not consider the moon explicitly.

Using results by \cite{Mat12} 300 planetesimals were distributed in a phase space region of the Hungaria group which has 
been shown to lead to an increased number of close encounters. Another 
648 were placed in the groups enclosing resonances.
The goal was to see how quickly the respective populations become so-called 
Near-Earth-Asteroids, where
every now and then one might have close encounters respectively impacts on the
Earth (and also Mars and Venus). The four different chosen regions, where the 
initial conditions for the four different samples were chosen, are given below:

\begin{itemize}

\item {\bf S1}:  300 Hungarias clones with three different semimajor axes 
a = 1.90792307, 1.91027822, 1.90508465 AU and equally distributed eccentricities in
the range $0.18 < e < 0.19$  and inclinations in the range of
$17^{\circ}<i<27^{\circ}$.

\item {\bf S2}:  216  clones close to the $\nu_{16}$ secular
  resonance\footnote{where the secular nodal motion of the massless body
    equals the nodal motion of Saturn} equally distributed with slightly
  larger semimajor axes than the Hungarias $1.9 < a < 2.1$ AU and the
  eccentricities and inclinations like in  {\bf S1}.

\item {\bf S 3}:  216  clones close to the $\nu_{5}$ secular
  resonance\footnote{where the secular perihelion motion of the massless body
    equals the perihelion motion of Jupiter} equally distributed with slightly
  smaller semimajor axes than the Hungarias $1.8 < a < 1.9$ AU and the
  eccentricities and inclinations like in {\bf S1}.
 
\item {\bf S4}:  216  clones in the region of the 
  $\nu_{5}$
secular resonance with semimajor axes $1.85 < a < 1.95$ AU, the eccentricities
like in the range of   {\bf S1} but with significantly larger inclinations
$27^{\circ}<i<35^{\circ}$.

\end{itemize}

\begin{figure}
\label{sec-res}
\includegraphics[height=.5\textheight]{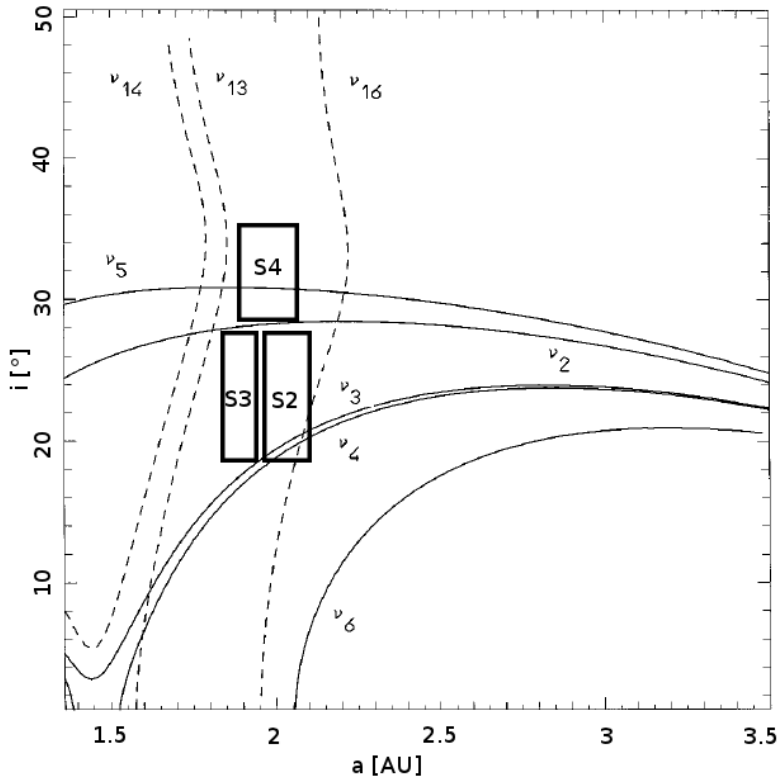}
\caption{The Hungaria asteroid region in a inclination versus semimajor axes
  diagramm. The locations of the initial conditions of the samples {\bf S2},
  {\bf S3} and {\bf S4} are shown in the rectangular boxes. The initial
  conditions for the sample {\bf S1}, just inside the Hungaria region, are
  between {\bf S2} and  {\bf S3}. The dashed lines indicate the secular
  resonances involving the longitudes, the solid lines involving the
   perihelion longitudes between a small body and a planet. The numbers '2' -
   '6' stand for the planets Venus to Saturn. (after \cite{Mic97}).}
\end{figure}

In Fig.\ref{sec-res} we depict the region of Hungaria family in an plot $\sin(i)$
versus the semimajor axes. Note that the bodies in samples {\bf S1},  {\bf S2}
and {\bf S3} have the same
inclinations but their initial conditions are shifted to larger respecively 
smaller semimajor axes. The
initial orbital elements for the fictitious bodies of {\bf S4} are distributed 
in semimajor axes $1.85 < a < 1.95$ and have large initial inclinations
(around $30^{\circ}$).
of the figure).

\subsection{Close Encounters with the Planets}

The results for the four different planetesimal samples are summarized in the 
following graphs 2-5. We note that during our integrations the mutual 
perturbations between planetesimals  was neglected and only
close encounters with the planets were reported.

\begin{figure}
\label{F1}
\includegraphics[height=.4\textheight]{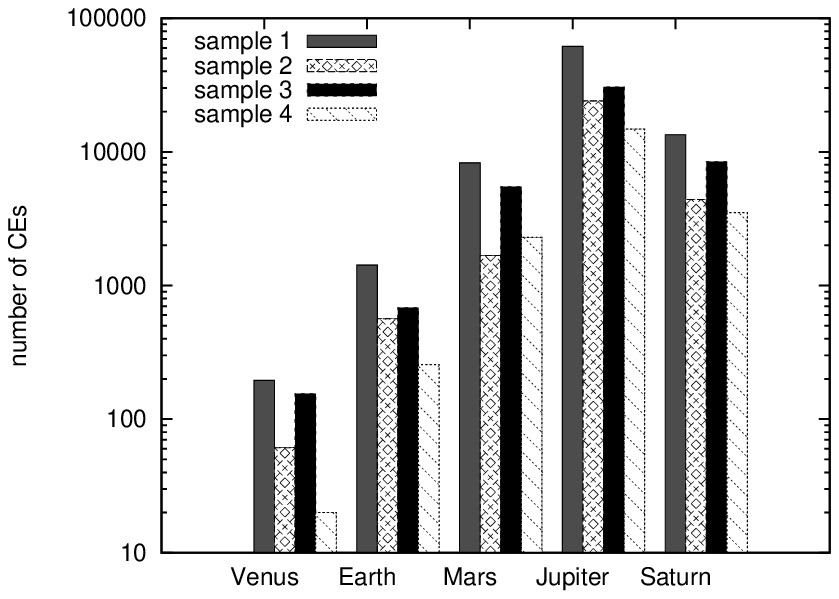}
\caption{Logarithmic plot of the number of close encounters 
of the fictitious objects with the planets
within its Hill's sphere (for more see in the text) 
for all the
planets involved. We separate the results for the four different samples} 
\end{figure}

Depending on the close encounters we could extrapolate collision timescales
which are crucial for estimates for a possible water transport onto the
terrestrial planets; we estimated the
water content to be three percent of the small bodies' masses.

\begin{figure}
\label{F3}
\includegraphics[height=.4\textheight]{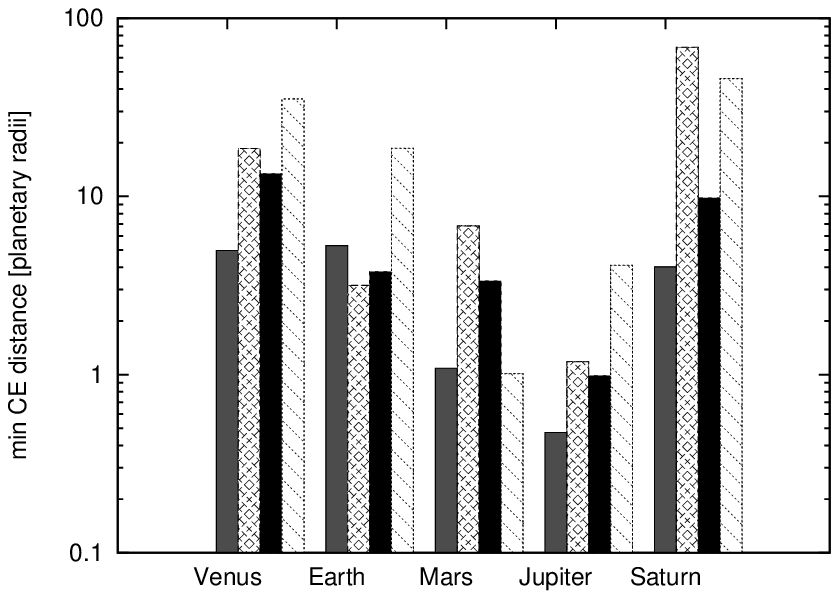}
\caption{Logarithmic plot of the closest encounters to the planets in units of
  the radii of the planets. Detailed description in the text}
\end{figure}

Fig.\ref{F1} shows the number of close encounters within the so-called Hill's
sphere\footnote{This sphere around a planet is defined as 
$r_H=(\frac{\mu}{3})^{\frac{1}{3}}$ 
where $\mu$ is the mass of the planet in Solar
masses. It can be regarded as a sphere of influence where inside the
gravitation of the planet is larger than the one of the Sun} .
We note that for Venus the results do not agree with other studies with
respect to the frequency of close encounters (e.g.\cite{Mat12},
\cite{Dvo99}). This is because of the relatively short integration time in our
investigations. The transport of the asteroids from the Hungaria region to the
inner regions of the planetary system takes longer (about several tenth of
million years) than for Mars and the Earth.
In Fig.\ref{F3} we compare the closest encounters during the integrations for the
four samples for all planets. One can see that counted in planetary radii only
one real collision occured and that is one with Jupiter. Although no
collisions are reported for the terrestrial planets we can extrapolate these
results of the frequency of the close encounters and find (see next chapter) an estimation of the time interval of a
single Hungaria clone for an encounter. It is also visible from the graph that
the (biased, see former remark about the integration time) tendency for
collisions is getting larger from Venus to Jupiter; this reflects 
the results shown in Fig.\ref{F1} where one can see the increasing number of close encounters from Venus to Jupiter. The larger values
for the closest distances to Saturn in planetary radii reflects also the smaller
number of encounters to this  planet.

\subsection{The Impacts}

\begin{table}
\caption{Impact times for the samples {\bf S1} -{\bf S4} (columns 3-6) onto the terrestrial
  planets within 1 Gyr} 
 \begin{tabular}{|c|c|c|c|c|c|}
 \hline
 Planet  & Radius[$AU^{-5}$]&  {\bf S1} &  {\bf S2}&  {\bf S3}&  {\bf S4}\\
 \hline
Mars & 2.25939 &  800.69 & 819.275 &  258.348 & 10.6991 \\
Earth & 4.25875 &  442.778 & 12.3894 & 30.3799 & 4.29435 \\
Venus & 4.04484 &  1.49954 & 71.7986 &  20.3137 &  0.655851\\
\hline
\end{tabular}
\label{tab-prob}
\end{table}

We need to say that in all our samples 'real' collisions were very rare! We
used the results of the many encounters to the planets to derive from there a
value for possible impacts (see Figs. 4 to 6). Binned values of the encounters
were plotted versus the number of such events. A logarithmic least square fit 
provided us with the desired value for the probability of collisions. We do
not show this fit for Venus because the results are biased because of the
small number of events. 

It is evident that Mars -- as
closest to the Hungarias -- suffers from impacts first of all, whereas Venus
globally is the planet with the least such events (due to the relatively short 
time scales of integrations). Most impacts of the fictitious objects occured
in {\bf S1}, which is a somewhat surprising fact, because shifting 
versus the secular resonances
({\bf S2} and {\bf S3}) should cause more perturbations on a body located there. Totally
insignificant for the transport of small bodies to the inner system seems to be
the group {\bf S4}, which is probably due to the large inclinations we have chosen
for the initial conditions. 

\begin{figure}
\label{F6}
\includegraphics[height=.4\textheight]{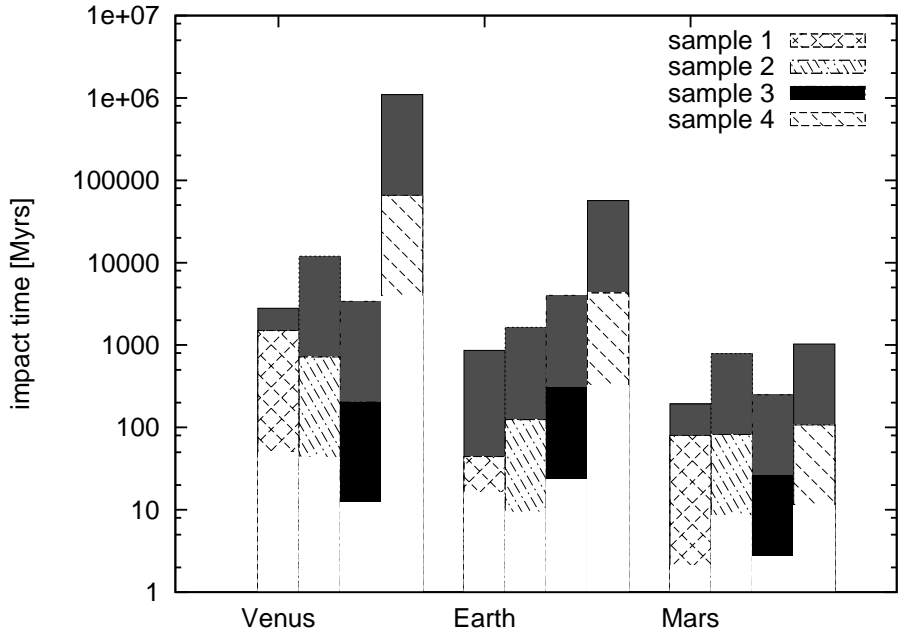}
\caption{Impact time scales for Hungaria like planetsimals in the
  samples {\bf S1} to  {\bf S4} on the terrestrial planets. For detail see
  text.
}
\end{figure}

In Fig.\ref{F6} we plotted the mean values of the impact time scales which are
the intersecting lines between the patterned and full bar segments. The
  mean values plus one standard deviation are denoted by the top of the bars,
  mean values minus one standard deviation by the bottom end of the patterned
  region. The large errors (especially for Venus and for all planets in {\bf S4} are
  caused by the poor statistics due to the choice of the integration time
  respectively the initial conditions.

In addition to the former results we have undertaken numerical experiments
with a fictitious planetary system consisting of more massive terrestrial
bodies comparable to a recent study by \cite{Sul07}. In our new study of the
sample {\bf S1}, also for 300 clones representing small bodies, we have
taken five times large masses\footnote{like in the former mentioned paper we
  define a multiplication factor $\kappa$ for the terrestrial planets} 
for these planets; we expected many more impacts
because of the higher gravitational perturbations. In fact in contrary to the
former results for the 'real' SS where only 1 'real' collision (namely with
Jupiter) was reported in this investigation the results are the following ones:

\begin{itemize}
\item 5 with Venus at 8.9, 8.58, 19.1, 34.7 and 38.6 myrs
\item 3 with Earth at 33.9, 43.13 and 46 myr
\item 2 with Mars 18.1 37.8 myr
\item 1 with Jupiter at 37.7 myr
\end{itemize}

These results agree much better with the ones we mentioned above, namely that
Venus is suffering the most of collisions. We can explain this -- expected
result -- that the time scales of transport of the 'planetesimals' to the 
inner SS are much faster in the case with $\kappa = 5$.

\begin{figure}
\label{F4}
\includegraphics[height=.4\textheight]{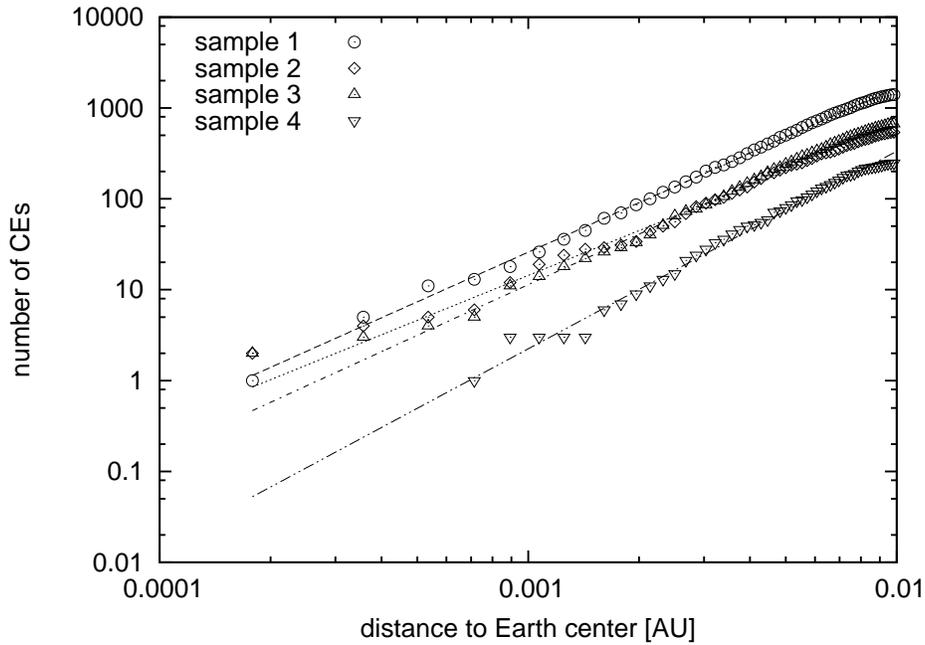}
\caption{Logarithmic least square fit for the encounters with Earth 
using the results of  samples {\bf S1} to  {\bf S4}}
\end{figure}

\begin{figure}\label{F5}
\includegraphics[height=.4\textheight]{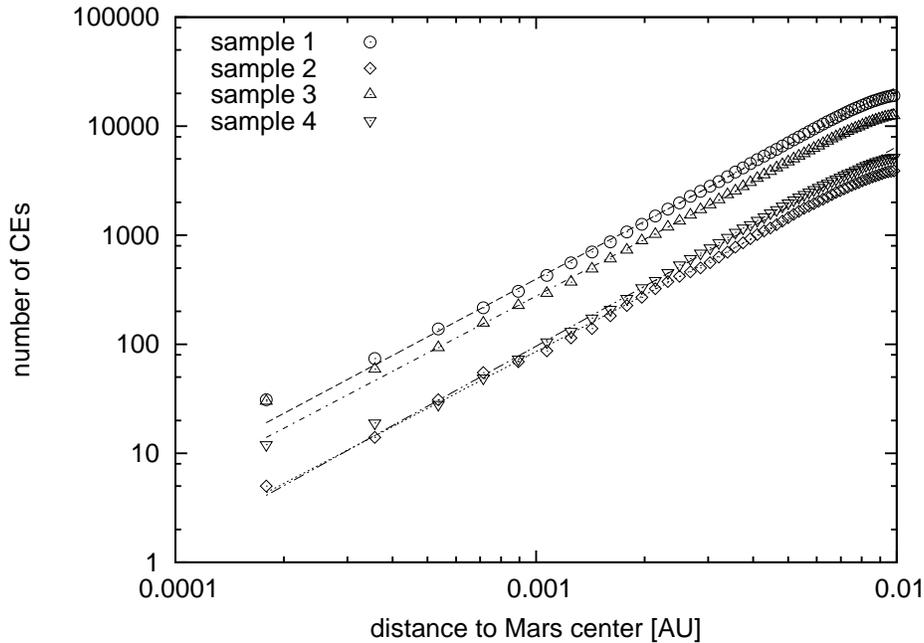}
\caption{Logarithmic least square fit for the encounters with Mars using the results of
  samples {\bf S1} to  {\bf S4}}
\end{figure}

\section{Conclusions: Water from Hungaria like planetesimals?}

The values from the former Tab.1 can now be used to estimate not only the how
many bodies from this region may hit the Earth, it can also be used to
estimate  -- in principle -- how much water was transported from this region
in the SS to our planet. Because of the very small number of 
impacts (see Tab.1) the
contribution to the water in the crust of the Earth (estimated to be  several $10^{-4}
M_{Earth}$) is for sure insignificant!

Another result is of interest in this context: by far Mars suffers from most
of such impacts and thus received a lot of more water even than Earth. But where is the
water now? New results show that man structures on the surface of Mars are due
to floating water. During the development of the Solar System this planet lost
most of its water because the thin atmosphere, the much lower gravitation field
and the absence of a protecting magnetosphere (\cite{Lam12}) 

For the water on Earth we can summarize that the phase space region 
around the Hungaria asteroid group is capable of injecting planetesimals 
into the inner Solar System but the total number is in fact far to low.
The timescales necessary for a considerable number of impacts  are too 
large to constitute an efficient water transport mechanism and to contribute
to our actual water on our planet and thus this region can be excluded as
source for water delivery to the Earth.

This preliminary study can be understood as a first step of investigation of
the whole phase space between Mars and Jupiter with respect to the transport
to terrestrial planet crossing regions (region of Near Earth Asterroids) and 
thus to possible collisions of small bodies with different water content on
these planets, especially on the Earth.




\begin{theacknowledgments}
The authors R.D., A.S., Z.S. and E.P.L. need to thank the NFN 
(Nationales Forschungsnetzwerk) 'Pathways to Habitable worlds'  from the
Fonds zur F\"orderung der Wissenschaft Nr. S 11603-N16 and S-11608-N16, S. Eggl would like to
acknowledge the support of University of Vienna's Forschungsstipendium 2012;
M.Galiazzo has to thank the  Doctoral School at the University of
Vienna 'From Asteroids to Impact Craters'

\end{theacknowledgments}

\bibliographystyle{aipprocl} 

\end{document}